\documentstyle[sprocl,epsfig,psfig,aps]{revtex}

\input{psfig}

\def\be{\begin{equation}}
\def\ee{\end{equation}}
\def\bea{\begin{eqnarray}}
\def\eea{\end{eqnarray}}

\newcommand{\newc}{\newcommand} 
\newc{\lra}{\leftrightarrow} 
\newc{\beq}{\begin{equation}} 
\newc{\eeq}{\end{equation}} 
\newc{\barr}{\begin{eqnarray}} 
\newc{\earr}{\end{eqnarray}} 

\begin{document} 
\title{SEARCHING FOR SUPERSYMMETRIC DARK MATTER. THE DIRECTIONAL RATE
AND THE MODULATION EFFECT DUE TO CAUSTIC RINGS . }

\author{J. D. VERGADOS} 

\address{Theoretical Physics Section, University of Ioannina, GR-45110, 
Greece\\E-mail:Vergados@cc.uoi.gr} 

\maketitle\abstracts{  The detection of the theoretically expected dark matter
is central to particle physics and cosmology. Current fashionable supersymmetric
models provide a natural dark matter candidate which is the lightest
supersymmetric particle (LSP). The allowed parameter space of such 
models combined with fairly well understood physics (quark substructure 
of the nucleon and nuclear structure) permit the evaluation of
the event rate for LSP-nucleus elastic scattering. The thus obtained 
event rates, which sensitively depend on the allowed parameter space 
parameters, are usually  very low or even undetectable.
 So, for background reduction, one would like to exploit two nice 
features of the reaction, the directional rate, which depends on the
sun's direction of motion and the modulation effect, 
i.e. the dependence of the event rate on  the earth's annual motion.
In the
present paper we study these phenomena in a specific class of non 
isothermal models,
which take into account the late in-fall of dark matter into our galaxy,
producing flows of caustic rings.
We find that the modulation effect arising from such models is smaller
than that found previously with isothermal symmetric velocity distributions
and much smaller compared to that obtained  
using a realistic asymmetric distribution
with enhanced dispersion in the galactocentric direction.
}
\section{Introduction}
In recent years the consideration of exotic dark matter has become necessary
in order to close the Universe \cite {KTP}$^,$ \cite{Jungm}. Furthermore in
in order to understand the large scale structure of the universe 
it has become necessary to consider matter
made up of particles which were 
non-relativistic at the time of freeze out. This is  the cold dark 
matter component (CDM). The COBE data ~\cite{COBE} suggest that CDM
is at least $60\%$ ~\cite {GAW}. On the other hand during the last few years
evidence has appeared from two different teams,
the High-z Supernova Search Team \cite {HSST} and the
Supernova Cosmology Project  ~\cite {SPF} $^,$~\cite {SCP} 
 which suggests that the Universe may be dominated by 
the  cosmological constant $\Lambda$r.
As a matter of fact recent data the situation can be adequately
described by  a barionic component $\Omega_B=0.1$ along with the exotic 
components $\Omega _{CDM}= 0.3$ and $\Omega _{\Lambda}= 0.6$.
In another analysis Turner \cite {Turner} gives 
$\Omega_{m}= \Omega _{CDM}+ \Omega _B=0.4$.
Since the non exotic component cannot exceed $40\%$ of the CDM 
~\cite{Jungm}$^,$~\cite {Benne}, there is room for the exotic WIMP's 
(Weakly  Interacting Massive Particles).
  In fact the DAMA experiment ~\cite {BERNA2} 
has claimed the observation of one signal in direct detection of a WIMP, which
with better statistics has subsequently been interpreted as a modulation signal
~\cite{BERNA1}.

The above developments are in line with particle physics considerations. Thus,
in the currently favored supersymmetric (SUSY)
extensions of the standard model, the most natural WIMP candidate is the LSP,
i.e. the lightest supersymmetric particle. In the most favored scenarios the
LSP can be simply described as a Majorana fermion, a linear 
combination of the neutral components of the gauginos and Higgsinos
\cite{Jungm}$^,$~\cite{JDV}$^-$\cite{Hab-Ka}. 

 Since this particle is expected to be very massive, $m_{\chi} \geq 30 GeV$, and
extremely non relativistic with average kinetic energy $T \leq 100 KeV$,
it can be directly detected ~\cite{JDV}$^-$\cite{KVprd} mainly via the recoiling
of a nucleus (A,Z) in the elastic scattering process:
\begin{equation}
\chi \, +\, (A,Z) \, \to \, \chi \,  + \, (A,Z)^* 
\end{equation}
($\chi$ denotes the LSP). In order to compute the event rate needs
the following ingredients:

1) An effective Lagrangian at the  
quark level in the context of supersymmetry as described 
in Refs.~\cite{Jungm}, Bottino {\it et al.} \cite{ref2} and \cite{Hab-Ka}.

2) A a quark model for the nucleon, needed in going from the quark to 
the nucleon level, since the obtained results are sensitive to the
presence of quarks other than u and d
~\cite{Dree,Adler,Chen}.

3) Compute the differential cross sections
using as reliable as possible many body nuclear wave functions.
~\cite{Ress}$^-$\cite{Nikol}.

The obtained rates sensitively depend on the input from  the allowed 
SUSY parameter space. 
 Since the expected rates are extremely low or even undetectable
with present techniques, one would like to exploit the characteristic
signatures provided by the reaction. Such are: a) The modulation
 effect, i.e the dependence of the event rate on the velocity of
the Earth and b) The directional event rate, which depends on the
 velocity of the sun around the galaxy as well as the the velocity
of the Earth. The latter effect, recognized sometime ago 
\cite {Sperg96}
has recently begun to appear feasible by the planned UKDMC experiment
\cite {UKDMC}.  We will study both of these effects in the present
work.

 In our previous letter
\cite{JDV99}and its subsequent expanded version \cite{JDV99b} we found
enhanced modulation, if one uses an appropriate asymmetric velocity
distribution with enhanced dispersion in the galactocentric direction
\cite{Druk}.

 The isolated galaxies are, however, surrounded by cold dark matter
, which,  due to gravity,  keeps falling continuously on 
them from all directions \cite{SIKIVIE}. As a result one has
caustic rings with matter density, which depends on space and
velocity.

 It is the purpose of our present paper to exploit the results of
such a scenario and calculate quantitavely the resulting 
modulation effect in the usual (differential and  total ) rate 
for LSP-nucleus elastic scattering. We will also study the
directional rates in the spirit of the recent work of Copi et al 
\cite{Copi99}.
For the reader's convenience we will give a very brief 
description of the basic
ingredients on how to calculate LSP-nucleus scattering cross section. 
We will
not, however, elaborate on how one gets the needed parameters from
supersymmetry. The calculation of these parameters  has become pretty 
standard.  One starts with   
representative input in the restricted SUSY parameter space as described in
the literature, e.g. Bottino {\it et al.} ~\cite{ref2}, 
Kane {\it et al.} , Castano {\it et al.} and Arnowitt {\it et al.} \cite {ref3}.

After this we will specialize our study in the case of the nucleus $^{127}I$, 
which is one of the most popular targets \cite{BERNA2}$^,$\cite{Smith}
\cite{Primack}$^,$.
We will present our results a function of the
LSP mass, $m_{\chi}$, in a way
which can be easily understood by the experimentalists.

\section{The Basic Ingredients for LSP Nucleus Scattering}
 Because of lack of space we are not going 
to elaborate here further on the construction of the effective Lagrangian
derived from supersymmetry, but refer the reader to the literature \cite
{JDV,KVprd,ref1,ref2,KVdubna}. For the reader's convenience we will
summarize our previous \cite {JDV99b} formulas related to the
LSP-Nucleus cross section and the event rates.
The effective Lagrangian can be obtained in first
order via Higgs exchange, s-quark exchange and Z-exchange. In 
a formalism familiar from the theory of weak interactions we write  
\beq
{\it L}_{eff} = - \frac {G_F}{\sqrt 2} \{({\bar \chi}_1 \gamma^{\lambda}
\gamma_5 \chi_1) J_{\lambda} + ({\bar \chi}_1 
 \chi_1) J\}
 \label{eq:eg 41}
\eeq
where
\beq
  J_{\lambda} =  {\bar N} \gamma_{\lambda} (f^0_V +f^1_V \tau_3
+ f^0_A\gamma_5 + f^1_A\gamma_5 \tau_3)N
 \label{eq:eg.42}
\eeq
and
\beq
J = {\bar N} (f^0_s +f^1_s \tau_3) N
 \label{eq:eg.45}
\eeq

We have neglected the uninteresting pseudoscalar and tensor
currents. Note that, due to the Majorana nature of the LSP, 
${\bar \chi_1} \gamma^{\lambda} \chi_1 =0$ (identically).
The parameters $f^0_V, f^1_V, f^0_A, f^1_A,f^0_S, f^1_S$  depend
on the SUSY model employed. In SUSY models derived from minimal SUGRA
the allowed parameter space is characterized at the GUT scale by five 
parameters,
two universal mass parameters, one for the scalars, $m_0$, and one for the
fermions, $m_{1/2}$, as well as the parameters 
$tan\beta $, one of $ A_0 $ (or  $ m^{pole}_t $)  and the
sign of $\mu $ \cite{ref3}. Deviations from universality at the GUT scale
have also been considered and found useful \cite{ref4}. We will not elaborate
further on this point since the above parameters involving universal masses
have already been computed in some models \cite{JDV,KVdubna} and effects
resulting from deviations from universality will be found elsewhere
(see Arnowitt {\it et al} in Ref. \cite{ref4} and 
Bottino  {\it et al} in Ref. \cite{ref2}). For some choices in the allowed
parameter space the obtained couplings can be found in a previous paper 
\cite{KVdubna}.

 With the above ingredients the differential cross section can be cast in the 
form 
\begin{equation}
d\sigma (u,\upsilon)= \frac{du}{2 (\mu _r b\upsilon )^2} [(\bar{\Sigma} _{S} 
            +\bar{\Sigma} _{V}~ \frac{\upsilon^2}{<\upsilon ^2>})~F^2(u)
                       +\bar{\Sigma} _{spin} F_{11}(u)]
\label{2.9}
\end{equation}
\begin{equation}
\bar{\Sigma} _{S} = \sigma_0 (\frac{\mu_r}{m_N})^2  \,
 \{ A^2 \, [ (f^0_S - f^1_S \frac{A-2 Z}{A})^2 \, ] \simeq \sigma^S_{p,\chi^0}
        A^2 (\frac{\mu_r}{\mu_r(N)})^2 
\label{2.10}
\end{equation}
\begin{equation}
\bar{\Sigma} _{spin}  =  \sigma^{spin}_{p,\chi^0}~\zeta_{spin}
\label{2.10a}
\end{equation}
\begin{equation}
\zeta_{spin}= \frac{(\mu_r/\mu_r(N))^2}{3(1+\frac{f^0_A}{f^1_A})^2}
[(\frac{f^0_A}{f^1_A} \Omega_0(0))^2 \frac{F_{00}(u)}{F_{11}(u)}
  +  2\frac{f^0_A}{ f^1_A} \Omega_0(0) \Omega_1(0)
\frac{F_{01}(u)}{F_{11}(u)}+  \Omega_1(0))^2  \, ] 
\label{2.10b}
\end{equation}
\begin{equation}
\bar{\Sigma} _{V}  =  \sigma^V_{p,\chi^0}~\zeta_V 
\label{2.10c}
\end{equation}
\begin{equation}
\zeta_V = \frac{(\mu_r/\mu_r(N))^2} {(1+\frac{f^1_V}{f^0_V})^2} A^2 \, 
(1-\frac{f^1_V}{f^0_V}~\frac{A-2 Z}{A})^2  \frac{<\upsilon ^2>} {c^2}  
[ 1  -\frac{1}{(2 \mu _r b)^2} \frac{2\eta +1}{(1+\eta)^2} 
\frac{\langle~2u~ \rangle}{\langle~\upsilon ^2~\rangle}] 
\label{2.10d}
\end{equation}
\\
$\sigma^i_{p,\chi^0}=$ proton cross-section,$i=S,spin,V$ given by:\\
$\sigma^S_{p,\chi^0}= \sigma_0 ~(f^0_S)^2$   (scalar) , 
(the isovector scalar is negligible, i.e. $\sigma_p^S=\sigma_n^S)$\\
$\sigma^{spin}_{p,\chi^0}= \sigma_0~~3~(f^0_A+f^1_A)^2$
  (spin) ,
$\sigma^{V}_{p,\chi^0}= \sigma_0~(f^0_V+f^1_V)^2$  
(vector)   \\
where $m_p$ is the proton mass,
 $\eta = m_x/m_N A$, $\mu_r$ is the LSP-nucleus reduced mass  
 , $\mu_r(N)$ is the LSP-nucleon reduced mass and  
\begin{equation}
\sigma_0 = \frac{1}{2\pi} (G_F m_N)^2 \simeq 0.77 \times 10^{-38}cm^2 
\label{2.7} 
\end{equation}
with
\begin{equation}
u = q^2b^2/2~~~or~~~
Q=Q_{0}u, \qquad Q_{0} = \frac{1}{A m_{N} b^2} 
\label{2.16b} 
\end{equation}
where
b is (the harmonic oscillator) size parameter, 
q is the momentum transfer  and
Q is the energy transfer to the nucleus\\
In the above expressions $F(u)$ is the nuclear form factor and
\begin{equation}
F_{\rho \rho^{\prime}}(u) =  \sum_{\lambda,\kappa}
\frac{\Omega^{(\lambda,\kappa)}_\rho( u)}{\Omega_\rho (0)} \,
\frac{\Omega^{(\lambda,\kappa)}_{\rho^{\prime}}( u)}
{\Omega_{\rho^{\prime}}(0)} 
, \qquad \rho, \rho^{\prime} = 0,1
\label{2.11} 
\end{equation}
are the spin form factors \cite{KVprd} ($\rho , \rho^{'}$ are isospin indices)
Both form factors are normalized to one at $u=0$.
$\Omega_0$ ($\Omega_1$) are the static isoscalar (isovector) spin 
matrix elements.

 The non-directional event rate is given by:
\begin{equation}
R=R_{non-dir} =\frac{dN}{dt} =\frac{\rho (0)}{m_{\chi}} \frac{m}{A m_N} 
\sigma (u,\upsilon) | {\boldmath \upsilon}|
\label{2.17} 
\end{equation}
 Where
 $\rho (0) = 0.3 GeV/cm^3$ is the LSP density in our vicinity and 
 m is the detector mass 
 
The differential non-directional  rate can be written as
\begin{equation}
dR=dR_{non-dir} = \frac{\rho (0)}{m_{\chi}} \frac{m}{A m_N} 
d\sigma (u,\upsilon) | {\boldmath \upsilon}|
\label{2.18}  
\end{equation}
where $d\sigma(u,\upsilon )$ was given above.

 The directional differential rate \cite{Copi99} in the direction
$\hat{e}$ is 
given by :
\begin{equation}
dR_{dir} = \frac{\rho (0)}{m_{\chi}} \frac{m}{A m_N} 
{\boldmath \upsilon}.\hat{e} H({\boldmath \upsilon}.\hat{e})
 ~\frac{1}{2 \pi}~  
d\sigma (u,\upsilon)
\label{2.20}  
\end{equation}
where H the Heaviside step function. The factor of $1/2 \pi$ is 
introduced, since  the differential cross section of the last equation
is the same with that entering the non-directional rate, i.e. after
an integration
over the azimuthal angle around the nuclear momentum has been performed.
In other words, crudely speaking, $1/(2 \pi)$ is the suppression factor we
 expect in the directional rate compared to the usual one. The precise 
suppression factor depends, of course, on the direction of observation.

\section{Convolution of the Event Rate}
 
 We have seen that the event rate for LSP-nucleus scattering depends on the
relative LSP-target velocity. In this section we will examine the consequences 
of the earth's
revolution around the sun (the effect of its rotation around its axis is
expected to
be negligible) i.e. the modulation effect. In the past this has been 
accomplished by
assuming a consistent LSP velocity dispersion, such as  
a Maxwell distribution ~\cite{Jungm} or asymmetric velocity distribution
with enhanced dispersion in the galactocentric direction
\cite{Druk,JDV99,JDV99b}.
More recently other very interesting non-isothermal
approaches have been proposed, which consider the in-fall of dark matter
into the galaxy producing flows of caustic rings. In particular the predictions of a self-similar model have been put forward as a possible scenario
for dark matter density-velocity distribution, see e.g. Sikivie et al
\cite{SIKIVIE}. The implications of such distributions on the
direct detection rates and, in particular, on the directional ones and 
the modulation effect are the subject of this work. Before proceeding
furtyher we should mention that after our manuscript
had been prepared another  approach has been suggested, which links the 
distribution to the density profile via Eddigton's formula 
\cite {Ullio}.

 Following Sikivie we will consider $2 \times N$ caustic rings, (i,n)
, i=(+.-) and n=1,2,...N (N=20 in the model of Sikivie et al),
each of which
contributes to the local density a fraction $\bar{\rho}_n$ of the
of the summed density $\bar{\rho}$ of each of the $i=+,-$. It contains
WIMP like particles with velocity 
${\bf y}^{'}_n=(y^{'}_{nx},y^{'}_{ny},y^{'}_{nz})$
in units of essentially
the sun's velocity ($\upsilon_0=220~Km/s$), with respect to the
galactic center.
The z-axis is chosen in the direction of the disc's rotation,
 i.e. in the direction of the motion of the
the sun, the y-axis is perpendicular to the plane of the galaxy and 
the x-axis is in the radial direction. We caution the reader that
these axes are traditionally indicated by astronomers as
$\hat{e}_{\phi},\hat{e}_r, \hat{e}_z)$ respectively. 
The needed quantities are taken from the 
work of Sikivie et al \cite{SIKIVIE} (see in TABLE I), via the 
definitions

$y^{'}_n=\upsilon_n/\upsilon_0
,y^{'}_{nz}=\upsilon_{n\phi}/\upsilon_0=y_{nz}
,y^{'}_{nx}=\upsilon_{nr}/\upsilon_0=y_{nx}
,y^{'}_{ny}=\upsilon_{nz}/\upsilon_0=y_{ny}
,\rho_{n}=d_n/\bar{\rho}
,\bar{\rho}=\sum_{n=1}^N~d_n$ and 
$y_n=[(y_{nz}-1)^2+y^2_{ny}+y^2_{nx}]^{1/2}$ (for each flow +.-).
 This leads to a
velocity distribution of the form:
\beq
f(\upsilon^{\prime}) = \sum_{n=1}^N~\delta({\bf \upsilon} ^{'}
    -\upsilon_0~{\bf y}^{'}_n)
\label{3.1}  
\eeq
Since the axis of the ecliptic \cite{KVprd} 
lies very close to the $y,z$ plane the velocity of the earth around the
sun is given by 
\beq
{\bf \upsilon}_E \, = \, {\bf \upsilon}_0 \, + \, {\bf \upsilon}_1 \, 
= \, {\bf \upsilon}_0 + \upsilon_1(\, sin{\alpha} \, {\bf \hat x}
-cos {\alpha} \, cos{\gamma} \, {\bf \hat y}
+cos {\alpha} \, sin{\gamma} \, {\bf \hat z} \,)
\label{3.6}  
\eeq
where $\alpha$ is the phase of the earth's orbital motion, $\alpha =2\pi 
(t-t_1)/T_E$, where $t_1$ is around second of June and $T_E =1 year$.

One can now express the above distribution in the laboratory frame 
\cite{JDV99b}
by writing $ {\bf \upsilon}^{'}={\bf \upsilon} \, + \, {\bf \upsilon}_E \,$ 


\section{Expressions for the Differential Event Rate.}

The mean value of the non-directional event rate of Eq. (\ref {2.18}), 
is given by
\beq
\Big<\frac{dR}{du}\Big> =\frac{\rho (0)}{m_{\chi}} 
\frac{m}{A m_N}  
\int f({\bf \upsilon}, {\boldmath \upsilon}_E) 
          | {\boldmath \upsilon}|
                       \frac{d\sigma (u,\upsilon )}{du} d^3 {\boldmath \upsilon} 
\label{3.10} 
\eeq
 The above expression can be more conveniently written as
\beq
\Big<\frac{dR}{du}\Big> =\frac{\rho (0)}{m_{\chi}} \frac{m}{Am_N} \sqrt{\langle
\upsilon^2\rangle } {\langle \frac{d\Sigma}{du}\rangle } 
\label{3.11}  
\eeq
where
\beq
\langle \frac{d\Sigma}{du}\rangle =\int
           \frac{   |{\boldmath \upsilon}|}
{\sqrt{ \langle \upsilon^2 \rangle}} f({\boldmath \upsilon}, 
         {\boldmath \upsilon}_E)
                       \frac{d\sigma (u,\upsilon )}{du} d^3 {\boldmath \upsilon}
\label{3.12}  
\eeq

 There are now experiments under way aiming at measuring directional 
rates \cite {UKDMC} using TPC counters which permit the observation
of the recoiling nucleus is observed in a certain direction.
From a theoretical point of view the directional rates have been
previously discussed by Spergel \cite {Sperg96} and Copi {\ et al}
\cite{Copi99}. 
The mean value of the directional differential event rate of Eq. (\ref {2.20}), 
is defined by
\beq
\Big<\frac{dR}{du}\Big>_{dir} =\frac{\rho (0)}{m_{\chi}} 
\frac{m}{A m_N} \frac{1}{2 \pi} 
\int f({\bf \upsilon}, {\boldmath \upsilon}_E)
{\boldmath \upsilon}.\hat{e} H({\boldmath \upsilon}.\hat{e})
                       \frac{d\sigma (u,\upsilon )}{du} d^3 {\boldmath \upsilon} 
\label{4.10} 
\eeq
where ${\bf \hat e}$ is the unit vector in the direction of observation. 
It can also be more conveniently expressed as
\beq
\Big<\frac{dR}{du}\Big>_{dir} =\frac{\rho (0)}{m_{\chi}} \frac{m}{Am_N} \sqrt{\langle
\upsilon^2\rangle } {\langle \frac{d\Sigma}{du}\rangle }_{dir} 
\label{4.11}  
\eeq
where
\beq
\langle \frac{d\Sigma}{du}\rangle _{dir}=\frac{1}{2 \pi} \int \frac{ 
{\boldmath \upsilon}.\hat{e} H({\boldmath \upsilon}.\hat{e})}
{\sqrt{ \langle \upsilon^2 \rangle}} f({\boldmath \upsilon}, {\boldmath \upsilon}_E)
                       \frac{d\sigma (u,\upsilon )}{du} d^3 {\boldmath \upsilon}
\label{4.12}  
\eeq
It is clear that the rate will depend on the direction of observation, 
showing a strong
correlation with the direction of the sun's motion. In a favorable 
situation the rate will merely be suppressed by a factor of $2 \pi$
relative to the non-directional rate. 
We will specialize our results in the case of caustic rings.
\subsection{The Non-directional Differential Event Rate in the Case
of caustic Rings}

Eq. \ref{3.12} takes the form
\beq
\langle \frac{d\Sigma}{du} \rangle  = \frac{2 \bar{\rho}}{\rho(0)}~
             a^2 [\bar {\Sigma} _{S} \bar {F}_0(u) +
        \frac{\langle \upsilon ^2 \rangle}{c^2}\bar {\Sigma} _{V} \bar {F}_1(u) 
                          +\bar {\Sigma} _{spin} \bar {F}_{spin}(u)]
\label{3.23}  
\eeq
We remind the reader that $\bar{\rho}$ was obtain for each type of flow
(+ or -), which explains  the factor of two. In the Sikivie model
\cite {SIKIVIE} we have $(2\bar{\rho}/\rho(0)\simeq1.0$, i.e. the whole dark
matter density lies in the form of caustic rings. In hybrid models,
which contain in addition an isothermal component, it is only a 
fraction, since the sum of all densities should be $\rho(0)$.

 The quantities
 $\bar{\Sigma} _{i},i=S,V,spin$ are given by Eqs. (\ref {2.10})-
(\ref {2.10c}). 
The quantities 
$\bar{F}_0,\bar{F}_1,\bar{F}_{spin}$  
are obtained from the corresponding form factors via the equations
\beq
\bar{F}_{k}(u) = F^2(u)\bar{\Psi}_k(u) \frac{(1+k) }{2k+1}~~,~~ k = 0,1
\label{3.24}  
\eeq
\beq
\bar{F}_{spin}(u) = F_{11}(u)\bar{\Psi}_0(u) 
\label{3.24a}  
\eeq
The functions $\tilde{\Psi}_k(u)$ depend on the model. 
Introducing the parameter
\beq
\delta = \frac{2 \upsilon_1 }{\upsilon_0}\, = \, 0.27,
\label{3.13}
\eeq
we find in the case of Sikivie model
\beq
\tilde{\Psi}_k(u)  =  \sqrt{\frac{2}{3}}~\sum_{n=1}^N~ \bar{\rho}_n
                   \tilde{y}_n^{2(k-1)}\Theta(\frac{\tilde{y}_n^2}{a^2}-u)
                       ~\tilde{y_n}
\label{3.26a}  
\eeq
with
\beq
a = \frac{1}{\sqrt{2} \mu _rb\upsilon _0}  
\label{3.27}  
\eeq
and
\beq
\tilde{y_n}= [(y_{nz}-1-\frac{\delta}{2}~sin\gamma~cos\alpha)^2
                 + (y_{ny}+\frac{\delta}{2}~cos\gamma~cos\alpha)^2
                 + (y_{nx}-\frac{\delta}{2}~sin\alpha)^2]^{1/2}
\label{3.28}  
\eeq
The above expressions (\ref{3.23})-(\ref{3.24a}) depend not only on $u$, but
on the phase of the Earth as well as in the angle $\gamma\simeq \pi/2$. If, 
however, in the $\Theta$-function we approxmate $\tilde{y}_n$ by $y_n$ given
by
\beq
y_n= [(y_{nz}-1))^2 + y_{ny}^2 + y_{nx}^2]^{1/2}
\label{3.29}  
\eeq
i.e. neglect threshold effects originating from the motion of the Earth,
the obtained expressions can be simplified. In fact, if we then expand 
$\tilde{\Psi}_k(u)$ in the small parameter $\delta$
and keep terms up to linear in it, the dependence on the phase of the
Earth can be seperated from the dependence on $u$.
Thus, to an approximation which is valid to no worse than $10\%$ for all u, 
the non-directional differential rate takes the form
\beq
\langle \frac{dR}{du} \rangle  = \bar{R} \frac{2 \bar{\rho}}{\rho(0)}
                   t~T(u) [1 - \cos \alpha~ H(u) ]
\label{3.31a}  
\eeq
where now $T(u)$ and $H(u)$ are now function of the variable $u$.

 The last expression must be compared to that of isothermal models:
\beq
\langle \frac{dR}{du} \rangle  = \bar{R} \frac{\rho^{'}(0)}{\rho(0)}
                   t~T(u) [1 + \cos \alpha~ H(u) ]
\label{3.31b}  
\eeq
with $\rho^{'}(0)$ being the density associated with the isothermal mode. 
In hybrid models it is less than $\rho(0)$.
In the above expressions $\bar{R}$ is the rate obtained in the conventional 
approach \cite {JDV} by neglecting the folding with the LSP velocity and the
momentum transfer dependence of the differential cross section, i.e. by
\beq
\bar{R} =\frac{\rho (0)}{m_{\chi}} \frac{m}{Am_N} \sqrt{\langle
v^2\rangle } [\bar{\Sigma}_{S}+ \bar{\Sigma} _{spin} + 
\frac{\langle \upsilon ^2 \rangle}{c^2} \bar{\Sigma} _{V}]
\label{3.39b}  
\eeq
where $\bar{\Sigma} _{i}, i=S,V,spin$ have been defined above, see Eqs
 (\ref{2.10}) - (\ref{2.10c}). 

The factor 
$T(u)$ takes care of the u-dependence of the unmodulated differential rate. It
is defined so that
\beq
 \int_{u_{min}}^{u_{max}} du T(u)=1.
\label{3.30a}  
\eeq
i.e. it is the relative differential rate. $u_{min}$ is determined by 
the energy cutoff due to the performance of the detector,i.e
\beq
u_{min}= \frac{Q_{min}}{Q_0}
\label{3.30c}  
\eeq
while $u_{max}$ is determined the via the relations:
\beq
u_{max}=min(\frac{y^2_{esc}}{a^2},max(\frac{y_{n} ^2}{a^2})~,~ n=1,2,...,N)
\label{3.30b}  
\eeq
On the other hand
$H(u)$ gives the energy transfer dependent
modulation amplitude (relative to the unmodulated amplitude).
The quantity $t$ takes care of the modification of the total rate due to the
nuclear form factor and the folding with the LSP velocity distribution.
 Since the functions $\bar{F}_0(u),\bar{F}_1$ and $\bar{F}_{spin}$ have 
a different dependence on u, the functions $T(u)$and $H(u)$ 
and $t$, in principle, depend somewhat
on the SUSY parameters. If, however, we ignore the small vector 
contribution and assume (i) the scalar and axial (spin) dependence on u is the
same, as seems to be the case for light systems \cite{DIVA00}$^,$\cite{Verg98},
 or (ii) only one
 mechanism (S, V, spin) dominates, the parameter $\bar{R}$ 
contains the dependence on all SUSY parameters. The parameters $t$ and
$T(u)$ depend on  the LSP mass and the nuclear parameters, while the 
$H(u)$ depends only on the parameter $a$. 
Eq. \ref{3.12} takes the form


\subsection{Expressions for the Directional Differential Event Rate}

The model of Sikivie et al \cite{SIKIVIE}, which is not spherically
 symmetric, offers itself as a perfect example for the study of the
directional rates. 
 Working as in the previous section we get \cite{JDV99b}
\beq
\langle \frac{d\Sigma}{du} \rangle_{dir}  = \frac{2 \bar{\rho}}
          {\rho(0)}~\frac{a^2}{2 \pi} [\bar {\Sigma} _{S} F_0(u) +
                     \frac{\langle \upsilon ^2 \rangle}{c^2}
                          \bar {\Sigma} _{V} F_1(u) +
                          \bar {\Sigma} _{spin} F_{spin}(u) ]
\label{4.39}  
\eeq
where the $\bar{\Sigma} _{i},i=S,V,spin$ are given by Eqs. (\ref {2.10})-
(\ref {2.10c}). 
The quantities 
$F_0,F_1,F_{spin}$ are obtained from the equations
\beq
F_k(u) = F^2(u)\Psi_k(u)\frac{(1+k)}{2k+1} , k = 0,1
\label{4.40}  
\eeq

\beq
F_{spin}(u) = F_{11}(u) \Psi_0(u) 
\label{4.41}  
\eeq
 In the Sikivie model we find
\barr
\Psi_k(u)& =&\sqrt{\frac{2}{3}}~\sum_{n=1}^N~ \bar{\rho}_n
                   \tilde{y}_n^{2(k-1)}\Theta(\frac{y_n^2}{a^2}-u)
   |(y_{nz}-1-\frac{\delta}{2}~sin\gamma~cos\alpha) {\bf e}_z.{\bf e}  
\nonumber\\
     &+&(y_{ny}+\frac{\delta}{2}~cos\gamma~cos\alpha){\bf e}_y.{\bf e}  
      +(y_{nx}-\frac{\delta}{2}~sin\alpha){\bf e}_x.{\bf e}|  
\label{4.42a}  
\earr
 In the model considered here the z-component of the LSP velocity,
with respect to the galactic center, for some rings  is smaller than
the sun's velocity, while for some others it
is larger. The components in the y and
the x directions are opposite for the + and - flows. So we will
distinguish the following cases: a) $\hat{e}$ has a
component in the sun's direction of
motion, i.e. $0<\theta < \pi /2$, labeled by u (up). b) Detection
in the direction specified by  $\pi /2 <\theta < \pi $, labeled by 
d (down).  The differential directional rate 
takes a different form depending on which quadrant
the observation is made. Thus, keeping terms up to linear in $\delta$,
we find:

1. In the first quadrant (azimuthal angle $0 \leq \phi \leq \pi/2)$.
\barr
\langle \frac{dR^i}{du} \rangle & = &\bar{R} \frac{2 \bar{\rho}}{\rho(0)}
    \frac{t}{2 \pi} T(u) [(R^i_z (u) - \cos \alpha~ H^i_1(u))
 |{\bf e}_z.{\bf e}|  
\nonumber \\ 
&+& (R^i_y +cos \alpha H^i_2 (u)+\frac{H^i_c (u)}{2}(|cos\alpha|+cos\alpha))
     |{\bf e}_y.{\bf e} | 
\nonumber \\ 
&+& (R^i_x -sin \alpha H^i_3 (u)+\frac{H^i_s (u)}{2}(|sin\alpha|-sin\alpha))
     |{\bf e}_x.{\bf e} | ]
\label{3.36a}  
\earr
2. In the second quadrant (azimuthal angle $\pi/2 \leq \phi \leq \pi)$
\barr
\langle \frac{dR^i}{du} \rangle & = &\bar{R} \frac{2 \bar{\rho}}{\rho(0)}
    \frac{t}{2 \pi} T(u) [(R^i_z (u) - \cos \alpha~ H^i_1(u)) 
  |{\bf e}_z.{\bf e}| 
\nonumber \\ 
&+& (R^i_y +cos \alpha H^i_2 (u)+\frac{H^i_c (u)}{2}(|cos\alpha|-cos\alpha))
     |{\bf e}_y.{\bf e} | 
\nonumber \\ 
&+& (R^i_x +sin \alpha H^i_3 (u)+\frac{H^i_s (u)}{2}(|sin\alpha|+sin\alpha))
     |{\bf e}_x.{\bf e} | ]
\label{3.37a}  
\earr
3. In the third quadrant (azimuthal angle $\pi \leq \phi \leq 3 \pi/2)$.
\barr
\langle \frac{dR^i}{du} \rangle & = &\bar{R} \frac{2 \bar{\rho}}{\rho(0)}
    \frac{t}{2 \pi} T(u) [(R^i_z (u) - \cos \alpha~ H^i_1(u)) 
  |{\bf e}_z.{\bf e}|  
\nonumber \\ 
&+& (R^i_y -cos \alpha H^i_2 (u)+\frac{H^i_c (u)}{2}(|cos\alpha|-cos\alpha))
     |{\bf e}_y.{\bf e} | 
\nonumber \\ 
&+& (R^i_x +sin \alpha H^i_3 (u)+\frac{H^i_s (u)}{2}(|sin\alpha|+sin\alpha))
     |{\bf e}_x.{\bf e} | ]
\label{3.38a}  
\earr
4. In the fourth quadrant (azimuthal angle $3 \pi/2 \leq \phi \leq 2 \pi)$
\barr
\langle \frac{dR^i}{du} \rangle & = &\bar{R} \frac{2 \bar{\rho}}{\rho(0)}
    \frac{t}{2 \pi} T(u) [(R^i_z (u) - \cos \alpha~ H^i_1(u)) 
       |{\bf e}_z.{\bf e}|  
\nonumber \\ 
&+& (R^i_y -cos \alpha H^i_2 (u)+\frac{H^i_c (u)}{2}(|cos\alpha|-cos\alpha))
     |{\bf e}_y.{\bf e} | 
\nonumber \\ 
&+& (R^i_x -sin \alpha H^i_3 (u)+\frac{H^i_s (u)}{2}(|sin\alpha|-sin\alpha))
     |{\bf e}_x.{\bf e} | ]
\label{3.39a}  
\earr
where $i=u,d$

By the reasoning given above, if one mechanism is dominant,
the parameters 
$R_x,R_y,R_z,H_1,H_2,H_3,H_c,H_s$ for both directions $u$ and $d$
depend only on $\mu_r$ and $a$.  They are all independent
 of the other SUSY parameters.


\section{The Total  Event Rates}

We will distinguish two possibilities, namely the directional and 
the non directional case. Integrating Eq. (\ref {3.31a}) in the
case of caustic rings we
obtain for the total non directional rate
\beq
R =  \bar{R}\, t \, \frac{2 \bar{\rho}}{\rho(0)}
          [1 - h(a,Q_{min})cos{\alpha})] 
\label{3.55}  
\eeq
 to be compared with the corresponding one for isothermal models:
\beq
R =  \bar{R}\, t \, \frac{\bar{\rho}^{'}(0)}{\rho(0)}
          [1 + h(a,Q_{min})cos{\alpha})] 
\eeq
In the above expressions  $Q_{min}$ is the energy transfer cutoff 
imposed by the detector.
 The modulation is described by the parameter $h$ only. 

 The effect of folding
with LSP velocity on the total rate is taken into account via the quantity
$t$. The SUSY parameters have been absorbed in $\bar{R}$. From our 
discussion in the case of differential rate it is clear that strictly
speaking the quantities $t$ and $h$ also depend on the SUSY parameters. They do 
not depend on them, however, if one considers the scalar, spin etc. modes 
separately. 

 Let us now examine the directional rate.  Integrating Eqs.
 (\ref {3.36a}) -  (\ref {3.39a}) we obtain:

1. In the first quadrant (azimuthal angle $0 \leq \phi \leq \pi/2)$.
\barr
R^i_{dir} & = &\bar{R} \frac{2 \bar{\rho}}{\rho(0)}
    \frac{t}{2 \pi} [(r^i_z  - \cos \alpha~ h^i_1) |{\bf e}_z.{\bf e}|  
\nonumber \\ 
&+& (r^i_y +cos \alpha h^i_2 +\frac{h^i_c }{2}(|cos\alpha|+cos\alpha))
     |{\bf e}_y.{\bf e} | 
\nonumber \\ 
&+& (r^i_x -sin \alpha h^i_3 +\frac{h^i_s }{2}(|sin\alpha|-sin\alpha))
     |{\bf e}_x.{\bf e} | ]
\label{3.56}  
\earr
2. In the second quadrant (azimuthal angle $\pi/2 \leq \phi \leq \pi)$
\barr
R^i_{dir} & = &\bar{R} \frac{2 \bar{\rho}}{\rho(0)}
    \frac{t}{2 \pi}  [(r^i_z  - \cos \alpha~ h^i_1) |{\bf e}_z.{\bf e}|  
\nonumber \\ 
&+& (r^i_y +cos \alpha h^i_2 (u)+\frac{h^i_c }{2}(|cos\alpha|-cos\alpha))
     |{\bf e}_y.{\bf e} | 
\nonumber \\ 
&+& (r^i_x +sin \alpha h^i_3 +\frac{h^i_s }{2}(|sin\alpha|+sin\alpha))
     |{\bf e}_x.{\bf e} | ]
\label{3.57}  
\earr
3. In the third quadrant (azimuthal angle $\pi \leq \phi \leq 3 \pi/2)$.
\barr
R^i_{dir} & = &\bar{R} \frac{2 \bar{\rho}}{\rho(0)}
    \frac{t}{2 \pi}  [(r^i_z  - \cos \alpha~ h^i_1) |{\bf e}_z.{\bf e}| 
\nonumber \\ 
&+& (r^i_y -cos \alpha h^i_2 (u)+\frac{h^i_c (u)}{2}(|cos\alpha|-cos\alpha))
     |{\bf e}_y.{\bf e} | 
\nonumber \\ 
&+& (r^i_x +sin \alpha H^i_3 +\frac{h^i_s }{2}(|sin\alpha|+sin\alpha))
     |{\bf e}_x.{\bf e} | ]
\label{3.58}  
\earr
4. In the fourth quadrant (azimuthal angle $3 \pi/2 \leq \phi \leq 2 \pi)$
\barr
R^i_{dir} & = &\bar{R} \frac{2 \bar{\rho}}{\rho(0)}
    \frac{t}{2 \pi}  [(r^i_z  - \cos \alpha~ h^i_1) |{\bf e}_z.{\bf e}|  
\nonumber \\ 
&+& (r^i_y -cos \alpha h^i_2 +\frac{h^i_c }{2}(|cos\alpha|-cos\alpha))
     |{\bf e}_y.{\bf e} | 
\nonumber \\ 
&+& (r^i_x -sin \alpha h^i_3 +\frac{h^i_s }{2}(|sin\alpha|-sin\alpha))
     |{\bf e}_x.{\bf e} | ]
\label{3.59 }  
\earr

\section{Discussion of the Results}

We have calculated the differential as well as the total event rates 
(directional and non directional) for elastic LSP-nucleus scattering 
using realistic nuclear form factors. We focused our attention on 
those aspects of the problem, which do not depend on the parameters of
supersymmetry other than the LSP mass.
 The parameter $\bar {R}$, normally calculated in SUSY theories, was not
calculated in this work. The interested reader is referred to the literature
\cite {ref1} $^,$ \cite {ref4} and, in our notation, to our previous work 
\cite {JDV} $^,$ \cite {KVprd} $^,$\cite {KVdubna}.\\

 We specialized our results for the target $^{127}I$.  We considered
the effects of the detector energy cutoffs, by studying two  typical
cases $Q_{min}=10,~20$ KeV.  Only the coherent mode
due to the scalar interaction was considered. The spin contribution will appear
elsewhere. 

Special attention was paid to the modulation effect
due to the annual motion of the earth. 
We assumed that the LSP density in our vicinity and the
velocity spectrum is that of caustic rings resulting from the self-
similar model of Sikivie et al \cite{SIKIVIE}.


 We will primarily concentrate on the total rates, which are 
described in terms of the quantities $t,r^i_x,r^i_y,r^i_z$ for the
unmodulated amplitude  and $h,h^i_1,h^i_2,h^i_3,h^i_c,h^i_s$ for the
modulated one. 
In TABLE II we show how these quantities vary with the detector energy cutoff
and the LSP mass. Of the above list only the quantities $t$ and $h$
enter the non directional rate. We notice that the usual
modulation amplitude $h$ is small. The main reason is that there are 
cancelations among the various
rings, since some rings are characterized by
$y_{nz}>1$, while for some others $y_{nz}<1$ (see TABLE I). Such
cancelations are less pronounced in the isothermal models.
 As expected, the parameter t, which contains the 
effect of the nuclear form factor and the LSP velocity dependence,
decreases as the reduced mass increases.

 We observe that
the quantities $h,r^i_j,h^i_j,~i=u,d$ and $j=x,y,z,c,s$, being the 
ratio of two amplitudes,
 are here essentially independent of the energy
cutoff $Q_{min}$. On the other hand the quantity $t$ decreases after
$Q_{min}$ is introduced, since an important part of the phase space is 
excluded.

 We notice that, unlike the isothermal models, the maximum of modulation
occurs around the 2nd of December, something already 
noticed by Sikivie et al \cite{SIKIVIE}. 

 Let us now examine the differential rates. We will begin with 
non-directional one, which is described in terms of the functions
$T(u)$ and $H(u)$. These are shown  for various
LSP masses and $Q_{min}$ in Fig. 1a for $T(u)$ and Fig. 1b for $H(u)$.
 We remind the reader that the dimensionless quantity u is related to the 
energy transfer Q 
via Eq. (\ref{2.17}) with $Q_0=60 KeV$ for $^{127}I$. 
Note that, due to our
normalization of T, the area under the corresponding curve is unity.
 This normalization was adopted to bring the various graphs on scale,
since the absolute values may vary 
substantially as a function of the reduced mass. We observe that the
function $T(u)$ differs somewhat from the predictions of the isothermal models
models. Here the function begins with a maximum at $u=0.0$, while in
the isothermal models the maximum occurs at $u=0.1$. Furthermore
this function shows less of a reduction as the reduced mass increases
(see Fig 1a). 
 The functions $H(u)$ exhibit step behavior in some
 regions. This is not unexpected, since the LSP velocity spectrum was
assumed to be discreet in the Sikivie model.

 The directional differential rates, which is now
beginning to look like a feasible experimental possibility
 \cite {UKDMC},is described by the functions\\
$~~~~~~~~~~~~R_x,R_y,R_z,H_1,H_2,H_3,H_c,H_s$\\
(for both directions $u$ and $d$)
 Due to lack of space we are not going to  show the here. We 
only mention that they
were defined as ratios of functions, with $T(u)$ in the
denominator. With this definition they are essentially independent of 
$u$, with values
approximately equal to their corresponding total values shown in TABLE II.


\section{Conclusions}
In the present paper we have calculated the parameters, which describe
the event rates for direct detection of supersymmetric dark matter.
We studied, in particular, the directional variation of the rates and the 
modulation  effect. 

The needed local density and velocity spectrum of the LSP were taken 
from the work of Sikivie et al \cite{SIKIVIE}, viewed as a late in-fall of dark
matter into our galaxy. They were derived in the context of a self-similar
model, which yields 40 caustic rings.

We presented our results in a suitable fashion, i.e by separating the
rates into two factors. One factor $\bar{R}$, which carries the
 dependence on the SUSY
parameters , not the subject of the present work, and another, which
is essentially independent of all SUSY parameters except the LSP mass. 
The latter depends mainly on the properties of the LSP velocity
distribution, the nuclear parameters and the kinematics. 
The nuclear form factor was taken into account and the effects of the detector
energy cut off were also considered.
Strictly speaking the obtained results describe the coherent process
in the case of $^{127}I$, but
we do not expect large changes, if the axial current is considered. 

 Our results, in particular the parameters $t$, see TABLE II, 
indicate that for large reduced mass, the kinematical advantage of
$\mu _r$ (see Eqs.
(\ref{2.10})- (\ref{2.10d}) is partly lost when the nuclear form factor and the 
convolution with the velocity distribution are taken into account. 
Also, if one  attempts to extract the LSP-nucleon cross section from
the data, in order to compare with the predictions of SUSY models,
one must take $t$ into account, since, for large reduced mass, $t$ is
different from unity.

In the case of the non-directional total event rates we find that 
the maximum no longer
occurs around June 2nd, but about six months later. The difference
between the maximum and the minimum is about $4\%$, a bit smaller
than that predicted by the symmetric isothermal models. It
is, however, substantially less than $h=0.46$ predicted by asymmetric
isothermal velocity distribution \cite{JDV99,JDV99b}. 

In the case of the directional rate we found that the rates depend on the 
direction of observation. The biggest rates are obtained, if the
observation is made close to the direction of the sun's motion.
The directional rates are suppressed compared to the usual 
non-directional rates by the factor 
$f_{red}=\kappa/(2 \pi)$. We find that $\kappa=r^u_z \simeq 0.7$, 
if the observation is made in the sun's direction of motion, while
$\kappa\simeq 0.3$ in the opposite direction.
The modulation is a bit larger than in the non-directional case, but the
largest value, 
$8\%$, is not obtained along the sun's direction of motion, but 
in the x-direction (galactocentric direction).
\section*{Acknowledgments}

The author would like to thank Professor Pierre Sikivie for bringing
to his attention the idea of caustic rings and velocity peaks. He would
like to 
acknowledge partial support of this work by 
TMR No  ERB FMAX-CT96-0090    
of the European Union. He would also like to thank the Humboldt
Foundation for their award that provided support during the final 
stages of this
work and Professor Faessler for his hospitality in Tuebingen. 

\newpage
\bigskip
\begin{table}[t]  
\caption{The velocity parameters  
$y^{'}_n =\upsilon_n/\upsilon_0,
~y_{nz} =y^{'}_{nz}=\upsilon_{n \phi}/\upsilon_0,
~y_{ny} =y^{'}_{ny}=\upsilon_{nz}/\upsilon_0,
~y_{nx} =y^{'}_{nx}=\upsilon_{nr}/\upsilon_0$ and
$y_n=[(y_{nz}-1)^2+y_{ny}^2+y_{nx}^2]^{1/2}$. Also shown are the quantities:
$a_n$, the caustic rind radii, and $\bar {\rho}_n=d_n/[\sum _{n=1}^{20} d_n]$. 
( For the other definitions see text ). 
}
\begin{center}
\footnotesize
\begin{tabular}{|l|c|rrrrrr|}
\hline
\hline
& & & & & &      \\
n &  $a_{n}(Kpc)$  & $y^{'}_n$  & $y_{nz}$ & $y_{ny}$  & $y_{nx}$ 
 & $y_n$ & $\bar {\rho}_n $\\
\hline 
& & & & & & &      \\
1 &38.0& 2.818& 0.636& $\pm$2.750& 0.000& 2.773 & 0.0120\\
2 &19.0& 2.568& 1.159& $\pm$2.295& 0.000& 2.301 & 0.0301\\
3 &13.0& 2.409& 1.591& $\pm$1.773& 0.000& 1.869 & 0.0601\\
4 & 9.7& 2.273& 2.000& $\pm$1.091& 0.000& 1.480 & 0.1895\\
5 & 7.8& 2.182& 2.000& 0.000& $\pm$0.863& 1.321 & 0.2767\\
6 & 6.5& 2.091& 1.614& 0.000& $\pm$1.341& 1.475 & 0.0872\\
7 & 5.6& 2.023& 1.318& 0.000& $\pm$1.500& 1.533 & 0.0571\\
8 & 4.9& 1.955& 1.136& 0.000& $\pm$1.591& 1.597 & 0.0421\\
9 & 4.4& 1.886& 0.977& 0.000& $\pm$1.614& 1.614 & 0.0331\\
10& 4.0& 1.818& 0.864& 0.000& $\pm$1.614& 1.619 & 0.0300\\
11& 3.6& 1.723& 0.773& 0.000& $\pm$1.614& 1.630 & 0.0271\\
12& 3.3& 1.723& 0.682& 0.000& $\pm$1.591& 1.622 & 0.0241\\
13& 3.1& 1.619& 0.614& 0.000& $\pm$1.568& 1.615 & 0.0211\\
14& 2.9& 1.636& 0.545& 0.000& $\pm$1.545& 1.611 & 0.0180\\
15& 2.7& 1.591& 0.500& 0.000& $\pm$1.500& 1.581 & 0.0180\\
16& 2.5& 1.545& 0.454& 0.000& $\pm$1.477& 1.575 & 0.0165\\
17& 2.4& 1.500& 0.409& 0.000& $\pm$1.454& 1.570 & 0.0150\\
18& 2.2& 1.455& 0.386& 0.000& $\pm$1.409& 1.537 & 0.0150\\
19& 2.1& 1.432& 0.364& 0.000& $\pm$1.386& 1.525 & 0.0135\\
20& 2.0& 1.409& 0.341& 0.000& $\pm$1.364& 1.515 & 0.0135\\
\hline
\hline
\end{tabular}
\end{center}
\end{table}
\begin{table}[t]  
\caption{The quantities $t$ and $h$ entering the total non directional
rate in the case of the
target $_{53}I^{127}$ for various LSP masses and $Q_{min}$ in KeV. 
Also shown are the quantities $r^i_j,h^i_j$
 $i=u,d$ and $j=x,y,z,c,s$, entering the directional rate for no energy
cutoff. For definitions see text. 
}
\begin{center}
\footnotesize
\begin{tabular}{|c|c|rrrrrrr|}
\hline
\hline
& & & & & & & &     \\
&  & \multicolumn{7}{|c|}{LSP \hspace {.2cm} mass \hspace {.2cm} in GeV}  \\ 
\hline 
& & & & & & & &     \\
Quantity &  $Q_{min}$  & 10  & 30  & 50  & 80 & 100 & 125 & 250   \\
\hline 
& & & & & & & &     \\
t      &0.0&1.451& 1.072& 0.751& 0.477& 0.379& 0.303& 0.173\\
h      &0.0&0.022& 0.023& 0.024& 0.025& 0.026& 0.026& 0.026\\
$r^u_z$&0.0&0.726& 0.737& 0.747& 0.757& 0.760& 0.761& 0.761\\
$r^u_y$&0.0&0.246& 0.231& 0.219& 0.211& 0.209& 0.208& 0.208\\
$r^u_x$&0.0&0.335& 0.351& 0.366& 0.377& 0.380& 0.381& 0.381\\
$h^u_z$&0.0&0.026& 0.027& 0.028& 0.029& 0.029& 0.030& 0.030\\
$h^u_y$&0.0&0.021& 0.021& 0.020& 0.020& 0.019& 0.019& 0.019\\
$h^u_x$&0.0&0.041& 0.044& 0.046& 0.048& 0.048& 0.049& 0.049\\
$h^u_c$&0.0&0.036& 0.038& 0.040& 0.041& 0.042& 0.042& 0.042\\
$h^u_s$&0.0&0.036& 0.024& 0.024& 0.023& 0.023& 0.022& 0.022\\
$r^d_z$&0.0&0.274& 0.263& 0.253& 0.243& 0.240& 0.239& 0.239\\
$r^d_y$&0.0&0.019& 0.011& 0.008& 0.007& 0.007& 0.007& 0.007\\
$r^d_x$&0.0&0.245& 0.243& 0.236& 0.227& 0.225& 0.223& 0.223\\
$h^d_z$&0.0&0.004& 0.004& 0.004& 0.004& 0.004& 0.004& 0.004\\
$h^d_y$&0.0&0.001& 0.000& 0.000& 0.000& 0.000& 0.000& 0.000\\
$h^d_x$&0.0&0.022& 0.021& 0.021& 0.020& 0.020& 0.020& 0.020\\
$h^d_c$&0.0&0.019& 0.018& 0.018& 0.017& 0.017& 0.017& 0.017\\
$h^d_s$&0.0&0.001& 0.001& 0.000& 0.000& 0.000& 0.000& 0.000\\
\hline 
& & & & & & & &     \\
t    &10.0&0.000& 0.226& 0.356& 0.265& 0.224& 0.172& 0.098\\
h    &10.0&0.000& 0.013& 0.023& 0.025& 0.025& 0.026& 0.026\\
\hline 
& & & & & & & &     \\
t    &20.0&0.000& 0.013& 0.126& 0.139& 0.116& 0.095& 0.054\\
h    &20.0&0.000& 0.005& 0.017& 0.024& 0.025& 0.026& 0.026\\
\hline
\hline
\end{tabular}
\end{center}
\end{table}
\setlength{\unitlength}{1mm}
\begin{figure}
\vspace*{1.8cm}
\begin{picture}(190,30)
\put(10,0){\epsfxsize=5cm \epsfbox{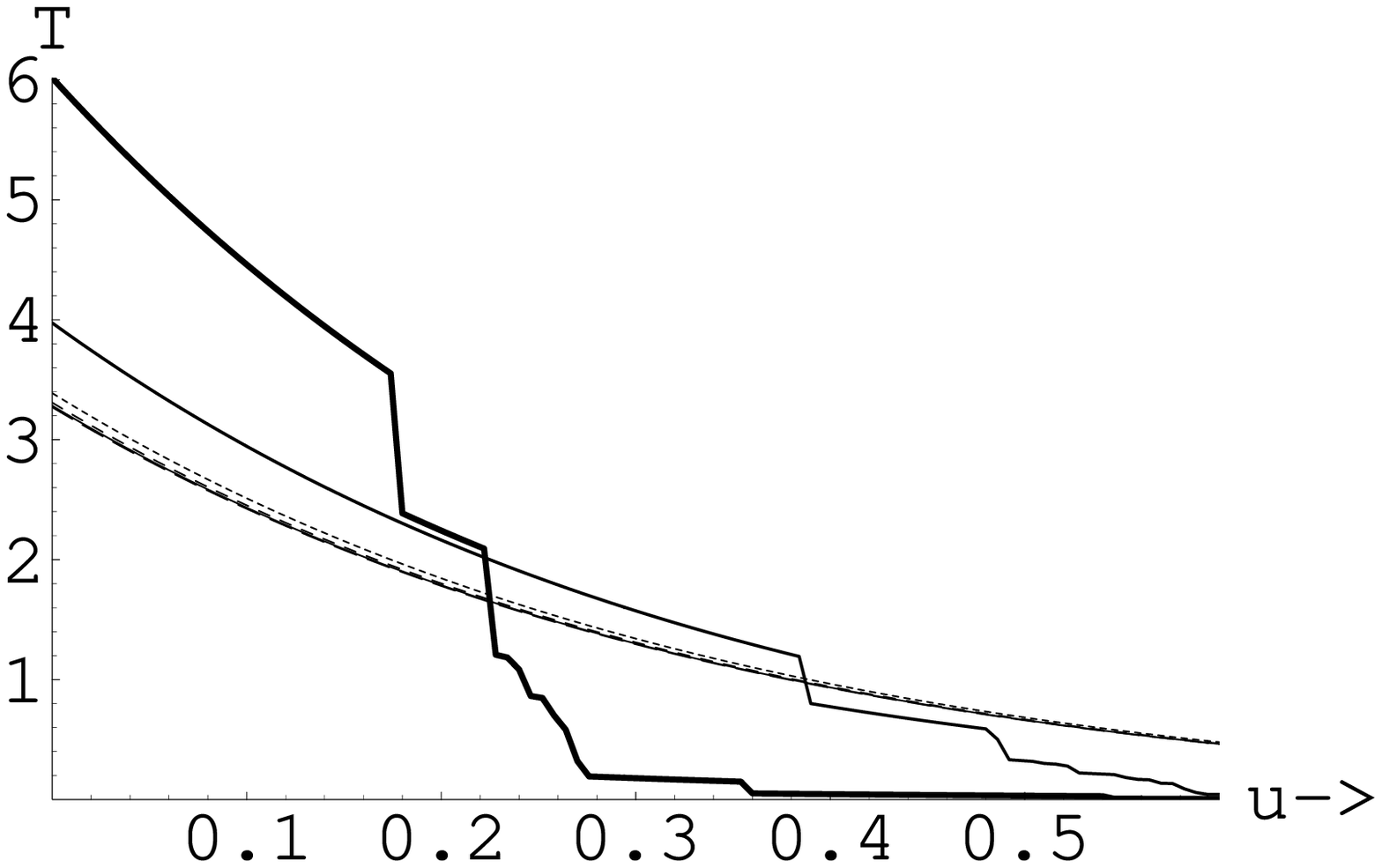}}
\put(70,0){\epsfxsize=5cm \epsfbox{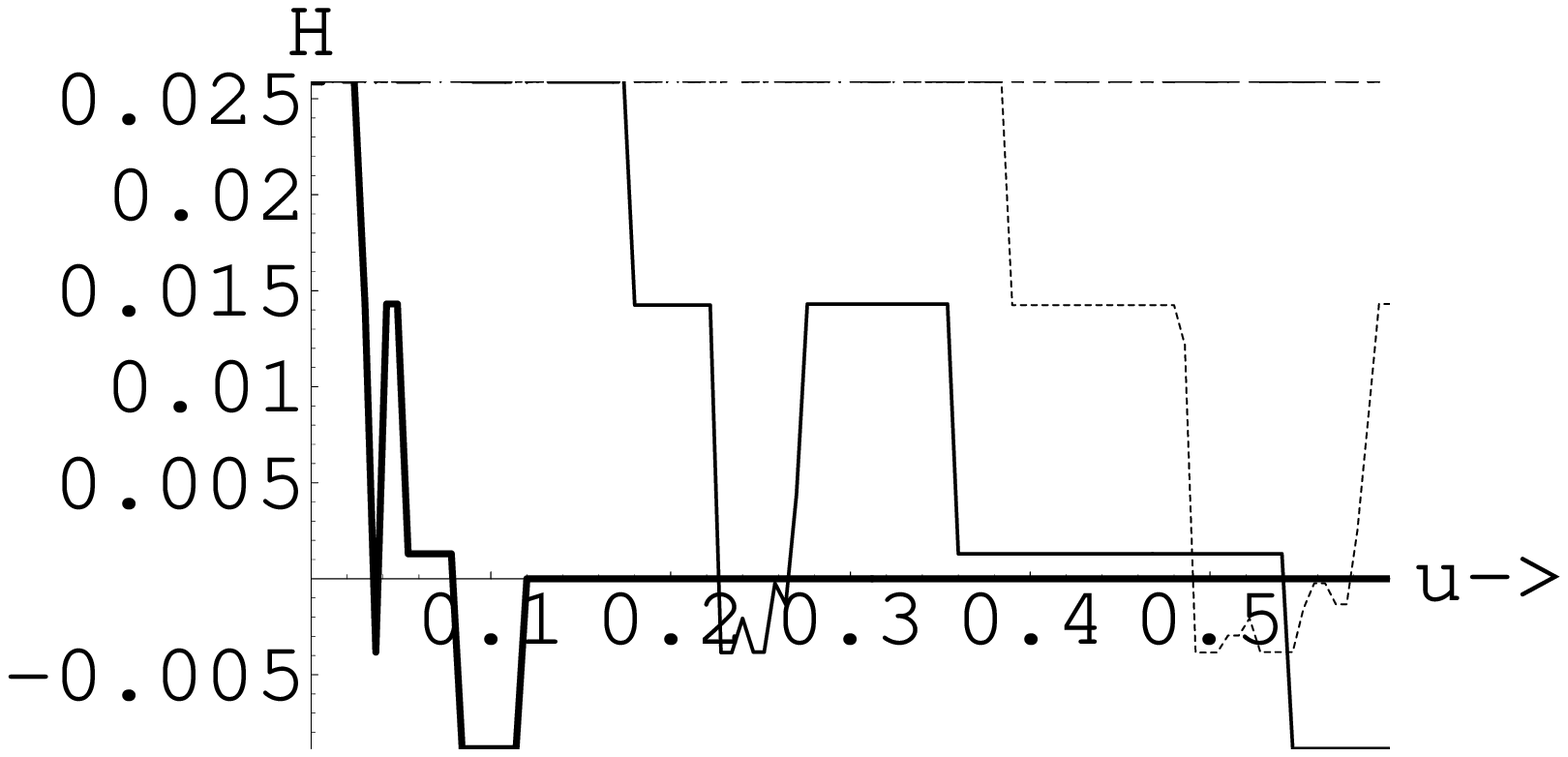}}
\end{picture}
\vspace*{-2.0cm}
\caption[]{The quantities $T(u)$ and $H(u)$ entering the differential
amplitude. Thick solid line corresponds to $m_{chi}=30~GeV$ the intermediate
thickness line to $m_{\chi}=80~ GeV$, the fine line to $m_{\chi}=100~GeV$.
The rest correspond to larger LSP masses and fall on top of each other.f
}
\label{fig.1}
\end{figure}

\section*{References}

\begin{thebibliography}{99}
 
\bibitem{KTP} E.W. Kolb and M.S. Turner, {\it The Early Universe}, Addison, 
Wesley, Redwood City, CA ,1990.\\
P.J.E. Peebles, {\it Principles of Physical Cosmology}, Princeton 
University Press, 1993.
M.S. Turner, {\it Cosmology:going beyond the big bang}, {\it Physics World},
 September 1996, p. 31.
\bibitem{Jungm}For a recent review see e.g.
G. Jungman {\it et al.},{\it  Phys. Rep.} {\bf 267}, 195  (1996).
\bibitem{COBE}G.F. Smoot et al., (COBE data), {\it Astrophys. J.} {\bf 396} 
(1992) L1.
\bibitem{GAW}E. Gawiser and J. Silk,{\it {Science}} {\bf 280}, 1405 (1988);
M.A.K. Gross, R.S. Somerville, J.R. Primack, J. Holtzman and A.A. Klypin,
{\it Mon. Not. R. Astron. Soc.} {\bf 301}, 81 (1998).
\bibitem{HSST}A.G. Riess {\it et al}, {\it Astron. J.}
{\bf 116} (1998), 1009.
\bibitem{SPF} R.S. Somerville, J.R. Primack and S.M. Faber, astro-ph/9806228;
{\it Mon. Not. R. Astron. Soc.} (in press).
\bibitem{SCP}Perlmutter, S. {\it et al} (1999) {\it Astrophys. J.}
{\bf 517},565; (1997) {\bf 483},565 ({\it astro-ph}/9812133).\\
S. Perlmutter, M.S. Turner and M. White, {\it Phys. Rev. Let.} {\bf 83},
670 (1999).
\bibitem{Turner} {\it Cosmological parameters},{\it astro-ph}/9904051; 
{\it Phys. Rep.} {\bf 333-334} (1990), 619.
\bibitem{Benne} D.P. Bennett {\it et al.}, (MACHO collaboration), A binary
lensing event toward the LMC: Observations and Dark Matter Implications, 
Proc. 5th Annual Maryland Conference, edited by S. Holt (1995);\\
C. Alcock {\it et al.}, (MACHO collaboration), {\it Phys. Rev. Lett.} {\bf 74}
, 2967 (1995). 
\bibitem{BERNA2} R. Bernabei et al., INFN/AE-98/34, (1998);
R. Bernabei et al., Phys. Lett. {\bf B 389}, 757 (1996).
\bibitem{BERNA1} R. Bernabei et al., Phys. Lett. {\bf B 424}, 195 (1998);
{\bf B 450}, 448 (1999).
\bibitem{JDV}J.D. Vergados, J. of Phys. {\bf G 22}, 253 (1996).
\bibitem{KVprd}T.S. Kosmas and J.D. Vergados, Phys. Rev. {\bf D 55}, 
1752 (1997).
\bibitem{Dree}M. Drees and M.M. Nojiri, Phys. Rev. {\bf D 48}, 3843 (1993);
Phys. Rev. {\bf D 47}, 4226 (1993). 
\bibitem{ref1}M.W. Goodman and E. Witten, Phys. Rev. {\bf D 31}, 3059 (1985);\\
K. Griest, Phys. Rev. Lett {\bf  61}, 666 (1988); 
Phys. Rev. {\bf D 38}, 2357 (1988) ; {\bf D 39}, 3802 (1989);\\
J. Ellis, and R.A. Flores, Phys. Lett. {\bf B 263}, 259 (1991);
Phys. Lett {\bf B 300}, 175 (1993); Nucl. Phys. {\bf B 400}, 25 (1993);\\
J. Ellis and L. Roszkowski, Phys. Lett. {\bf B 283}, 252 (1992).
\bibitem{JDV99} J.D. Vergados, Phys. Rev. Let. {\bf 83}, 3597 (1999)
\bibitem{JDV99b} J.D. Vergados, Phys. Rev. {\bf D 62}, 0235XX-1 (2000);
 astro-ph/0001190
\bibitem{ref2}A. Bottino {\it et al.}, Mod. Phys. Lett. {\bf A 7}, 733 (1992);
Phys. Lett. {\bf B 265}, 57 (1991); Phys. Lett {\bf B 402}, 113 (1997);
hep-ph $/$ 9709222; hep-ph $/$9710296;\\
J. Edsjo and P Gondolo, Phys. Rev. {\bf D 56}, 1789 (1997);\\
Z. Berezinsky {\it et al.}, Astroparticle Phys. {\bf 5}, 1 (1996);\\
V.A. Bednyakov, H.V. Klapdor-Kleingrothaus and S.G. Kovalenko,
 Phys. Lett.  {\bf B 329}, 5 (1994).
\bibitem{ref3}G.L. Kane {\it et al.}, Phys. Rev. {\bf D 49}, 6173 (1994);\\
D.J. Casta\v no, E.J. Piard and P. Ramond, Phys. Rev. 
{\bf D 49}, 4882 (1994); \\ D.J. Casta\v no, Private Communication;\\ 
A.H. Chamseddine, R. Arnowitt and P. Nath, Phys. Rev. Lett. {\bf 49},
970 (1982);
P. Nath, R. Arnowitt and A.H. Chamseddine, Nucl. Phys. {\bf B 227}, 121 (1983);
R. Arnowitt and P. Nath, Mod. Phys. Lett. {\bf 10}, 1215 (1995);
R. Arnowitt and P. Nath, Phys. Rev. Lett.  {\bf 74}, 4952 (1995);
R. Arnowitt and P. Nath, Phys. Rev. {\bf D 54}, 2394 (1996).
\bibitem{ref4} R. Arnowitt and P. Nath, hep-ph$/9701301$; hep-ph/9902237;\\
S.K. Soni and H.A. Weldon, Phys. Lett. {\bf B 126}, 215 (1983);\\
V.S. Kapunovsky and J. Louis, Phys. Lett. {\bf B 306}, 268 (1993);\\
M. Drees, Phys. Lett {\bf B 181}, 279 (1986);\\
P. Nath and R. Arnowitt, Phys. Rev. {\bf D 39}, 279 (1989);\\
J.S. Hagelin and S. Kelly, Nucl. Phys. {\bf B 342}, 95 (1990);\\
Y. Kamamura, H. Murayama and M. Yamaguchi, Phys. Lett. {\bf B 324}, 52 (1994);\\
S. Dimopoulos and H. Georgi, Nucl. Phys. {\bf B 206}, 387 (1981).
\bibitem{Hab-Ka}H.E. Haber and G.L. Kane, Phys. Rep. {\bf  117} 75 (1985).
\bibitem{Adler}S.L. Addler, Phys. Rev. {\bf D 11} (1975) 3309.
\bibitem{Chen}T.P. Cheng, Phys. Rev. {\bf D 38} 2869 (1988);
H -Y. Cheng, Phys. Lett. {\bf B 219} 347 (1989).
\bibitem{Ress}M.T. Ressell {\it et al.}, Phys. Rev. {\bf D 48}, 5519 (1993);
M.T. Ressell and D.J. Dean, hep-ph$/9702290$.
\bibitem{Dimit}V.I. Dimitrov, J. Engel and S. Pittel, 
Phys. Rev. {\bf D 51}, R291 (1995).
\bibitem{Engel}J. Engel, Phys. Lett. {\bf B 264}, 114 (1991).
\bibitem{Nikol}M.A. Nikolaev and H.V. Klapdor-Kleingrothaus, Z. Phys. 
{\bf A 345}, 373 (1993); Phys. Lett {\bf 329 B}, 5 (1993); Phys. Rev. {\bf
D 50}, 7128 (1995).
\bibitem{Sperg96} D. Spergel, {\it Phys. Rev.} {\bf D 37}, 299 (1996).
\bibitem{UKDMC}K.N. Buckland, M.J. Lehner, G.E. Masek, in {\it Proc.
1st Int. Workshop on IDM}, p. 475, Ed. N.J.C. Spooner,
 World Scientific, (1997); Proc. {\it 3nd Int. Conf. on Dark Matter
in Astro- and part. Phys.} (Dark2000), Ed. H.V. Klapdor-Kleingrothaus,
Springer Verlag (2000).
\bibitem{Copi99}Copi, C.J., Heo, J. and Krauss, L.M. (1999),
\bibitem{Druk} A.K. Drukier et al., Phys. Rev. {\bf D 33}, 3495 (1986);\\
J.I. Collar et al., Phys. Lett {\bf B 275}, 181 (1992).
\bibitem{SIKIVIE}P. Sikivie, I. Tkachev and Y. Wang Phys. Rev. Let. {\bf 75}, 
2911 (1995; Phys. Rev. {\bf D 56}, 1863 (1997)\\
P. Sikivie, Phys. Let. {\bf b 432}, 139 (1998); astro-ph/9810286 
\bibitem{Smith}P.F. Smith et al, Phys. Rep. {\bf 307}, 275 (1999); N. Spooner,
               Phys.Rep. {\bf 307}, 253 (1999);
J.J. Quenby et al Astropart. Phys. {\bf 5}, 249 (1996); Phys. Lett. {\bf 351},
                                   70 (1995).
\bibitem{Primack}J.R. Primack, D. Seckel and B. Sadoulet, Ann. Rev. Nucl.
Part. Sci. {\bf 38}, 751 (1988).
\bibitem{Smith1}P.F. Smith and J.D. Lewin, Phys. Rep. {\bf 187}, 203 (1990). 
\bibitem{KVdubna}J.D. Vergados and T.S. Kosmas, Physics of Atomic nuclei,
 Vol. 61, No 7, 1066 (1998) (from Yadernaya Fisika, Vol. 61, No 7, 1166 (1998).
\bibitem{Ullio}P. Ullio and M. Kamioinkowski, Velocity distributions and
annual-modulation signatures of weakly interacting massive particles,
hep-ph/0006183.
\bibitem{DIVA00}P.C. Divari, T.S. Kosmas, J.D. Vergados and L.D. Skouras,
Shell Model Calculations for LSP Scattering Off Light Nuclei, to appear in 
Phys. Rev C.
\bibitem{Verg98}J.D. Vergados, Phys. Rev. {\bf D 58}, 103001-1 (1998);
\end{thebibliography}
\end{document}